# Towards a Flexible Intra-Trustcenter Management Protocol

Vangelis Karatsiolis[1,2], Marcus Lippert[1], Alexander Wiesmaier[1], Anna Pitaev[1], Markus Ruppert[1], Johannes Buchmann[1]

[1] Technische Universität Darmstadt,
Department of Computer Science,
Hochschulstraße 10, D-64289 Darmstadt, Germany
`karatsio,mal,wiesmaie,epchtana,mruppert@cdc.informatik.tu-darmstadt.de`
[2] Darmstadt Centre of IT Security,
Hochschulstraße 10, D-64289 Darmstadt, Germany
`karatsio@cdc.informatik.tu-darmstadt.de`

**Abstract.** This paper proposes the Intra Trustcenter Protocol (ITP), a flexible and secure management protocol for communication between arbitrary trustcenter components. Unlike other existing protocols (like PKCS#7, CMP or XKMS) ITP focuses on the communication within a trustcenter. It is powerful enough for transferring complex messages which are machine and human readable and easy to understand. In addition it includes an extension mechanism to be prepared for future developments.

## 1 Introduction

Public Key Infrastructure (PKI) is used to provide security of communication in electronic processes like Internet and e-commerce. Most implementations and specifications (e.g. [RFC3280], [RFC2510], [RFC2511]) consider hierarchical PKIs where trust into the authenticity of all keys depends on a Certification Authority (CA) and sometimes on an optional Registration Authority (RA). But one can easily think of more authorities in the PKI and trustcenter environment.

Trustcenter software has to provide a great variety of services each carried out by complex processes like registration, key generation, certification, personalisation of personal security environments, key backup and many more. Each service has different requirements on the operational environment. The requirements may even alter when the trustcenter has to adapt to different situations. A common technique in software engineering to meet the demands of an adaptive software is to break it up into components, each covering certain operational aspects. These components can then be distributed according to the needs of the special environment. Even more, certain components can be replaced to reflect changes in the environment and so on. Clearly, a component based approach to complex processes needs well suited means of communication among these components. Furthermore, this communication is a grand issue to the overall security of the trustcenter.

## 1.1 Contribution

In this paper we propose a new protocol which is well suited for the communication between arbitrary trustcenter components. The proposed protocol takes a look into an area which current standards do not address, namely the communication between arbitrary trustcenter components. Section 2 names already existing protocols and discusses their properties. Secondly, we list the requirements for intra-trustcenter communication which serve as design criteria for our protocol. This is discussed in Section 3. Moreover we motivate the use of XML for specifying the syntax of our protocol, since XML is a well accepted means for specifying data structures. XML comes with sophisticated security mechanisms which we employed to guarantee integrity, authenticity and confidentiality as well as enforce operational security constraints as roles or dual control. Furthermore, the syntax of the protocol is given along with an example from a trustcenter scenario. These can be found in Section 4. In Section 5 the protocol is analysed regarding its flexibility and security along with its deployment considerations. Lastly, Section 6 gives an outlook on further developments of the protocol.

## 2 Certificate and Certification Protocols

The existing proposals and standards concerning certificate and certification management can be divided into three groups.

The first one consists of the "Cryptographic Message Syntax Standard" [PKCS7] and "Certification Request Syntax Standard" [PKCS10]. They were developed by RSA Laboratories and describe message formats in ASN.1 [X.680]. PKCS#10 is only suitable for an end-entity to provide a to be certified public key to a trustcenter and thus is not general enough for intra-trustcenter communication. PKCS#7 comes in two modes: In the more general one it can be used as a container for transporting arbitrary data in a secure way. This of course can also be employed within a trustcenter. But it has the disadvantage that it does not consider the transported content, leading therefore to coarser security mechanisms. In the special enrollment mode it is only usable for delivering certificates and certificate revocation lists and thus not general enough.

The second group contains two protocol standards published by the Internet Engineering Task Force (IETF). These are the CMP [RFC2510] and the CMC [RFC2797] specifications. They, in turn, are based on the CRMF [RFC2511] and CMS [RFC2630] formats which play a similar role as PKCS#10 and PKCS#7, respectively. They are all specified in ASN.1. These protocols focus on scenarios where communication takes place between an end-entity, a certification authority and optionally a registration authority. Thus they can not be employed for the communication between arbitrary components.

A rather new development is the "XML Key Management Specification" [XKMS] which was proposed by the W3C and arose from the need for a key management protocol for the XML signature [XMLSig] and encryption [XMLEnc] standards. It is not appropriate for intra-trustcenter communication since it focuses on the communication between end-entities and trustcenters.

Only the CMP and PKCS#7 specifications can be actually considered related work to our protocol, since they are used for the purpose of communication inside a trustcenter. However, as we already pointed out there are some considerations using these protocols inside the trustcenter and our approach offers an alternative to these two specifications.

## 3 Protocol Criteria

This section focuses on the qualities a communication protocol between trustcenter components should have. The criteria that should apply to the protocol itself (how should it be) are named, along with a discussion on the motivation behind them.

### 3.1 Design criteria

This subsection shows the criteria that must apply to a protocol for intra-trustcenter communication. This list is partly based on the design goals of other protocols (e.g [RFC2510], [XKMS]) and partly on what we have encountered while setting up trustcenters for different projects.

1. *Generality* The protocol must be able to handle all kind of data which may be passed between trustcenter components. It must not put any restrictions upon the number and type of components the trustcenter is composed of. The protocol should allow to express the data in a structured manner when possible (i.e. not just binary). [3]
   As the security and flexibility mechanisms of the protocol have to apply to all messages, it is clear that all possible messages must be presentable within the protocol. In order to be able to apply different mechanisms[4] to different parts of the message it must be well structured.

2. *Extensibility* The protocol must allow to be extended by new messages, data types or components.
   This is necessary to be able to meet future requirements, like new trustcenter products or special request types.

3. *Independency* The protocol has to be independent from the structure and workflow of the trustcenter. This includes the independence of any means of transportation, including transport media, online versus offline transport, security and protection issues.
   This is to meet all possible connection types for the trustcenter components and all policies which might be applied.

---

[3] signatures will surely be binary data

[4] e.g. encryption of parts of the message

4. *Automation* The protocol must allow the automation of trustcenter processes.

   The trustcenter may automate some processes. The protocol must not prohibit this. This includes especially the authentication mechanisms.

5. *Scalability* The protocol must scale with the size of the public key infrastructure, the security level and the complexity of the trustcenter installation.

   Different trustcenters have different complexity, size and security level. In addition these parameters may change in the lifetime of a trustcenter. The protocol has to deal with these facts.

6. *Traceability* The protocol has to allow the tracing of messages, applications and products. The messages have to be human readable where possible.[5]

   As trustcenters are highly security sensitive applications there will be a lot of auditing and quality insurance means applied to them. The protocol must support this. Additionally this eases the debugging or error search.

7. *Security* The protocol messages (or parts of it) have to be able to be secured (i.e. authentic, unaltered and secret). Additionally the protocol must support common authorisation techniques (e.g. dual control, delegation). It must be possible to achieve this with any cryptographic algorithm or data structure, regarding both the communicated data and the protection of the protocol messages itself.

   As the intra-trustcenter communication is a major issue in the security of the whole trustcenter the protocol must support all means for protecting this communication. As the field of cryptography develops, the protocol must be able to adjust the means of protection to the current state of the art.

## 4 The ITP Protocol

### 4.1 Why to use XML?

XML is ideal for describing structured data. Applications using XML can define a special syntax in order to describe the data that should be used inside the application. For example a special tag `<X509Certificate>` can describe an X.509 certificate while the tag `<revocationPassword>` describes the revocation password that an end-user would like to use. This flexible data format is perfectly suited for defining the structures of ITP.

This data format is in addition portable. Data described in XML can be exchanged between applications which are XML aware and almost all programming languages support XML. Introducing document type definition (DTD) or XML schema, the tags and structure of the XML document are constrained to the rules applied from the DTD or schema. This can be used to avoid errors in the XML structure.

XML documents are both human and machine readable. This is a significant property of an XML document in cases where humans should examine the correctness of data. For example the subject distinguished name (DN) that should be placed in the certificate can be such data. This can be described in the `<subjectDN>` tag (e.g. `<subjectDN>CN=Alice Cryptographer, O=MyOrg, C=DE</subjectDN>`). While a machine can examine if this tag contains correct data it can not examine whether the name Alice Cryptographer really exists. A human administrator on the contrary can examine the credentials of the entity and decide whether such a value for this tag is valid or not.

---

[5] encrypted parts will surely be binary data

XML supports also digital signatures [XMLSig] and encryption [XMLEnc]. Moreover it supports different cryptographic mechanisms (like RSA, DSA or ECDSA in the future) providing therefore security related to different mathematical problems (factoring and discrete logarithm problem). Another possibility that the signature framework has to offer is that XML structures can be multiply signed from one, two or more different entities (this can include also physical persons) as well as ability to sign certain portions of the whole structure. For example in the case of dual control the presence of two signatures is required.

Lastly, using XML someone can prevent the use of ASN.1 structures for exchanging such messages. Applications can avoid the step of implementing these special and in several cases complicated structures just to serve the communication between trustcenter components. The description of the ITP messages in XML is quite simple to implement and use and does not require special software (for example a library for ASN.1).

### 4.2 ITP messages

ITP consists of XML messages exchanged between components inside the trustcenter. The most significant element is an `<application>` which encapsulates all data that is exchanged between such components.

**Application** An `<application>` represents a request for a trustcenter component to deliver a certain service.[6] In principal, every component can form an `<application>` and head it to any other component. The basic idea is the following: In order to have a trustcenter execute a certain task, one or more structures of type `<application>` are generated. Each `<application>` is routed from component to component. Each component performs its service and, if necessary, modifies the `<application>` by adding, changing or removing parts of it. Thus, an `<application>` contains the pieces of data needed to accomplish a task as well as it determines how this data should be processed. The syntax of an `<application>` can be found in Figure 1.

An `<application>` mandatorily owns the attribute `id` which uniquely identifies it. If the same application travels among different components this `id` always remains unchanged. By this traceability of applications can be achieved. Furthermore, replay of already processed applications can be prevented. This is of great importance since a trustcenter may have to guarantee that certain services are

---

[6] we have chosen the name `<application>` instead of `<request>` since every request is actually a form (like an application form) from which every component extracts information that it needs and fills the form with the information it produces

```
<message version="1.0" id="123456789">

    <sender>Registration</sender>

    <recipient>Certification</recipient>

    <application id="12345678">
        <profile id="ProfileId">
            <element1Name>value</element1Name>
            <element2Name>value</element2Name>
            <!--  ... more elements  -->
        </profile>
        <ds:Signature> <!--  signature elements...  --> </ds:Signature>
        <!--  ... more signatures  -->
    </application>

    <!--  ..... more applications  -->

    <ds:Signature> <!--  signature elements...  --> </ds:Signature>
    <!--  ..... more signatures  -->

</message>
```

**Fig. 1.** Syntax of an ITP message

carried out at most once per request (e.g. a request for just one certificate should under no circumstances lead to more than one certificate being issued).

The tag `<profile>` determines by its unique attribute `id` the kind of application. The most basic ones concern issuing certificates and processing revocations, but other may be added as well. It is also possible to provide `<profile>`-ids for communicating meta information within the trustcenter.

The `<profile>` also defines, how the applications are processed and which components are involved at which stage of procession. This, among others, includes whether a key-pair is to be created in the trustcenter or provided by the end-entity, possibly along with some proof-of-possession data, or if some kind of key backup has to be done. Since it is left to the components, how to accomplish their services in detail, it is also they who have to interpret the `<profile>`-ids. This mechanism enables ITP to transport arbitrary requests.

Within the `<profile>`-tag the pieces of data are included. Most prominent among them are certificates and CRLs. Apart from this other data can be met like passwords, end-entity data or trustcenter functional data. For example this could be the last and first name of a person for an application to registration or the subject DN for an application to the certification component. In any case a way to express this data is needed. This is a special XML tag that describes each of them. The usual data exchanged among the components possess their own tag but it is possible to define new tags that represent data which are needed in order for an application to be complete.

**Application signature** An `<application>` may contain one or more elements `<ds:Signature>` of type enveloped signature according to the XML signature standard as described in [XMLSig]). These signatures are calculated over the `<application>` by the last component which altered its content. By this, the changes are authorised. The destination component which imports the application may verify the signature of the application. In addition, it is possible for components to sign certain portions of an `<application>`. XML signatures allow signing of a given path inside an XML structure. This gives a component the ability to sign only the data it creates and actually is responsible for. In addition with the possibility to calculate and append more that one signature over an `<application>` the last can travel among the trustcenter components, which now verify the signature only over data that they need to operate. Furthermore dual control can be enabled with two participants signing the XML structure appending both signatures at the `<application>` (one participant can be a trustcenter administrator).

We see that with the use of XML signatures, as well as the ability to define a new `<application>` and express all possible data inside a trustcenter we achieve the two goals of ITP. That is security and flexibility.

**Message** The element that is root of every ITP message is the `<message>`. A `<message>` element contains two attributes. The first one is the `version` attribute to denote the version of ITP in use. Current version is "1.0". The second one is the `id` attribute. The `id` attribute holds a unique number among messages for each message. Therefore each component can recognise whether it has already processed this message or not. In addition a `<message>` must contain the `<sender>` and `<recipient>` elements.

The `<sender>` element describes the component sending the `<application>`. This can be an IP address or even a symbolic name known inside the trustcenter. Name resolving based on an IP is for example ideal in an environment in which the trustcenter components operate in different hosts and based on the symbolic name in the case where the components are found in the same host. The same rule applies to the `<recipient>` element used to hold information about the component to which the message is destined for.

Inside the trustcenter a peer-to-peer communication may exist, where every component can communicate with another component. For example, the CA should be able to receive messages from different RAs, send messages to a backup component as well as to a Directory Services component. We see that this protocol is flexible on the naming of every peer and the problem of defining the source and the destination of the messages is solved.

A `<message>` is a container for `<application>` elements. A `<message>` may contain one or more `<application>` elements. For example the registration component sends three certification requests at the same time. One can be a request for an encryption key-pair, the second for a signature and the third for a non-repudiation. In this case it must envelope three `<application>` elements inside the `<message>`. Of course a new `<application>` (see also 4.3) can be defined, which represents exactly this use case and therefore only one `<application>` is needed, demonstrating the flexibility of the proposed scheme. A complete ITP message can be seen in Figure 1.

A `<message>` can also be secured with an XML signature. The same properties as in the signing of an `<application>` element apply to this case.

### 4.3 An example of the protocol

In order to demonstrate the use of the protocol we construct the following scenario. This scenario consists of a Registration and a Certification component (which is offline) as well as a Directory Services component. All components have multi-client capability and they can operate for different virtually hosted CAs. In addition, two trustcenter administrator (operators) must sign all request coming to the offline Certification component.

Alice, who belongs to organisation A, wants to get three certificates. One for encryption, one for digital signature and one for non-repudiation. In addition she provides a revocation password to be used in case of revocation. She also wants her certificates to be publicly available.

Alice makes a certification request to the Registration component which examines the credentials of Alice. It checks that they are valid and sends a message to Certification component for the special certification request (known to the system with the id "MultiCert"). This message contains the subject DN for the certificate, the virtual CA that it should sign the certificates, the hash value of the revocation password, information whether the certificates should be published or not and the e-mail address of Alice. It then writes the message with one application and signs the application to provide authenticity. Lastly, two trustcenter operators sing the message too, in order to be accepted from the Certification component. The message can be seen in Figure 2. This message is written in a portable media and is manually transported to the Certification component (since it is offline) from the two operators.

The Certification component imports the message, verifies the signatures of the operators over the application and logs their identities with the subject DN of each operator certificate.[7] It then verifies the signature of the Registration component, creates all three certificates (signed with the key of the appropriate virtual CA) and changes and adds information to the application into a new message to the Directory Services. It also signs the application to provide authenticity and integrity of the data. The key used for signing should not be the same as the one used for signing the certificates. In addition, the signatures in the messages exchanged between components are operational signatures without any special properties. The message (seen in Figure 3) is transfered to the Directory Services.

---

[7] such a requirement may appear when it comes to evaluation of the Certification component

```
<message version="1.0" id="20040202164445">
  <sender>Registration</sender>
  <recipient>Certification</recipient>
  <application id="20040202164832">
   <profile id="MultiCert">
    <clientName>Host A</clientName>
    <subjectDN>CN=Alice,OU=OrgUnitName,O=OrgName,C=DE</subjectDN>
    <revocationPassword>7c4a8 ... 8941c</revocationPassword>
    <email>alice@orgunitname.orgname.de</email>
    <publiclyAvailable>true</publiclyAvailable>
   </profile>
   <ds:Signature> other signature elements ....  </ds:Signature>
   <ds:Signature> signature of first operator   </ds:Signature>
   <ds:Signature> signature of second operator  </ds:Signature>
  </application>
</message>
```

**Fig. 2.** Registration to Certification message

Lastly, the Directory Services component verifies the signature. Afterwards it examines whether the certificates should be published or not (specified in the `<publiclyAvailable>` tag), it publishes them and sends Alice a notification by e-mail with the certificates attached.

In this scenario two messages were created. One from Registration to Certification and one from Certification to Directory Services. Both refer to the same application. Therefore the `id` of the application must remain the same. Another remark is that the `<subjectDN>` field of the first message is missing in the second message. This field is of great importance for the Certification component, since if this field is missing it can not issue the certificates, but of no importance for the Directory Services component. In addition the subject's DN is now contained in the certificates.

## 5  Analysis of ITP

ITP aims at a secure and flexible protocol for the communication between trust-center components. The mechanisms that ITP has to offer enable these two goals inside the trustcenter.

### 5.1  Flexibility

ITP messages can travel among all trustcenter components. The sender and recipient tags of a message resolve the addressing problem. Every component is in position to set these tags according to the desired flow of information. In addition, the number of the operating components can vary without affecting the protocol itself.

```
<message version="1.0" id="20040202170134">
  <sender>Certification</sender>
  <recipient>Directory Services</recipient>
  <application id="20040202164832">
   <profile id="MultiCert">
      <clientName>Host A</clientName>
      <encCertificate>Base64 encoded certificate</encCertificate>
      <signCertificate>Base64 encoded certificate</signCertificate>
      <nonRepCertificate>Base64 encoded certificate</nonRepCertificate>
      <revocationPassword>7c4a8 ... 8941</revocationPassword>
      <email>alice@orgunitname.orgname.de</email>
      <publiclyAvailable>true</publiclyAvailable>
   </profile>
   <ds:Signature> other signature elements .... </ds:Signature>
  </application>
</message>
```

**Fig. 3.** Certification to Directory Services message

All kind of data inside the trustcenter can be transported with ITP. Both binary data (like the X.509 certificate) and simple text can be send along with an ITP message.

There is the possibility to define new tags and applications. Therefore, ITP is in position to meet new requirements in the trustcenter communication. This may occur in cases where the communicated data or the services offered from the trustcenter change.

ITP messages are XML messages and thus can be send with various mechanisms. These may include HTTP(S), TCP/IP, email or even a floppy disc. In addition, the messages themselves are portable among applications since they are described in XML. Moreover they can be read from humans as well as machines.

The messages and applications can be tracked. Every message and application possesses a unique id. This enables the components or the administrators to keep a history of past messages or applications. In addition, the flow of information inside the trustcenter can be followed. This is especially useful in cases where auditing or special logging is required.

### 5.2 Security

ITP offers integrity, authenticity and confidentiality to the communication of a trustcenter's components. It further supports authorisation mechanisms.

Messages can be secured for confidentiality. The whole message or portions of it may be encrypted. Candidates for such an action may be values like the private key of an end-entity, the revocation password or even the postal address of the end-user.

It is possible to calculate a signature over a message, an application or certain data inside an application. This makes it possible to provide authenticity as well as integrity for these values.

ITP messages may be signed from one or more components or administrators. This enables dual or multi control. This kind of control is met very often in high level security trustcenters.

### 5.3 Deploying ITP

ITP, as described in this paper, can be used from trustcenters which seek a method for arranging their information flow inside the trustcenter, independently of the number and nature of components and services as well as the kind of information communicated among them. This is a significant step, since the communication between arbitrary trustcenter components with the requirements mentioned in Section 3 is difficult to achieve and existent solutions do not provide this functionality (instead of, most of them are standardised). At the moment ITP can be used only for intra-trustcenter specific implementations. A newer version of ITP is needed that specifies the ITP messages, in order for them to describe most of the tasks inside a trustcenter in a standardised way that will enable interoperability of different implementations.

## 6   Future Work

We proposed ITP, an XML-based protocol for intra-trustcenter communication. We introduced the design criteria of our protocol, argued on their usefulness and reasoned on whether these criteria have been met or not. We specified the 1.0 version of ITP along with an example extracted from a real scenario.

But a lot of work still needs to be done. Our next goal is to design and implement ITP version 2.0, a fully-fledged protocol with extended functionality and strictly defined rules and structure (work towards a standardised version). We further want to investigate the use of criticality flags for the various tags of the protocol messages. Towards this direction considerations on whether all messages should default to non-critical or which actions should be taken if a criticality rule is violated should be made. A mechanism for error handling should also be integrated in the protocol with clearly defined exceptions (like signature verification failed or application has already been processed) and exception identifiers (ids). In addition a more advanced signing policy mechanism can be integrated. With this mechanism it can be regulated whether components are allowed to change data from the initial application and therefore destroy the original signature and its validity. Special care should be taken on examining the trade-offs between simplicity and extended functionality.

## References

RFC2510.  C. Adams, S. Farrell, *Internet X.509 Public Key Infrastructure Certificate Management Protocols*, Request for Comments 2510, March 1999.

AL99. C. Adams, S. Lloyd, *Understanding Public-Key Infrastructure: Concepts, Standards and Deployment Considerations*, New Riders Publishing, 1999.

RFC2630. R. Housley, *Cryptographic Message Syntax*, Request for Comments 2630, June 1999.

HP01. R. Housley, T. Polk, *Planning for PKI*, John Wiley & Sons, Inc., 2001.

RFC3280. R. Housley, W. Polk, W. Ford, D. Solo, *Internet X.509 Public Key Infrastructure Certificate and Certificate Revocation List (CRL) Profile*, Request for Comments 3280, April 2002.

RFC2511. M. Myers, C. Adams, D. Solo D. Kemp, *Internet X.509 Certificate Request Message Format*, Request for Comments 2511, March 1999.

RFC2797. M. Myers, X. Liu, J. Schaad, J. Weinstein, *Certificate Management Messages over CMS*, Request for Comments 2797, April 2000.

RFC2314. B. Kaliski, *PKCS #10: Certification Request Syntax Version 1.5*, Request for Comments 2314, March 1998.

RFC2315. B. Kaliski, *PKCS #7: Cryptographic Message Syntax Version 1.5*, Request for Comments 2315, March 1998.

PKCS7. RSA Laboratories, *PKCS#7 v1.5: Cryptographic Message Syntax Standard*, November 1993.

PKCS10. RSA Laboratories, *PKCS#10 v1.7: Certification Request Syntax Standard*, May 2000.

PKCS12. RSA Laboratories, *PKCS#12 v1.0: Personal Information Exchange Syntax*, June 1999.

XML. *Extensible Markup Language (XML) 1.0 (Second Edition)*,
`http://www.w3.org/TR/REC-xml` (24 Jan. 2004).

XKMS. *XML Key Management Specification (XKMS)*,
`http://www.w3.org/TR/2001/NOTE-xkms-20010330/` (9 Feb. 2004).

XMLEnc. *XML Encryption Syntax and Processing*,
`http://www.w3.org/TR/xmlenc-core/` (9 Feb. 2004).

XMLSig. *XML-Signature Syntax and Processing*,
`http://www.w3.org/TR/xmldsig-core/` (24 Jan. 2004).

X.509. ITU-T Recommendation X.509, *Information Technology - Open Systems Interconnection - The Directory: Authentication Framework*, August 1997.

X.680. ITU-T Recommendation X.680, *Information Technology - Abstract Syntax Notation One (ASN.1): Specification of basic notation*, July 2002.

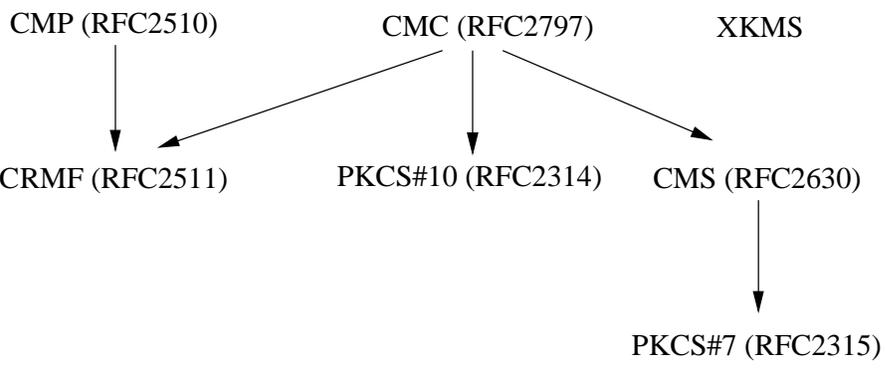

CMP (RFC2510)          CMC (RFC2797)          XKMS

CRMF (RFC2511)    PKCS#10 (RFC2314)    CMS (RFC2630)

PKCS#7 (RFC2315)

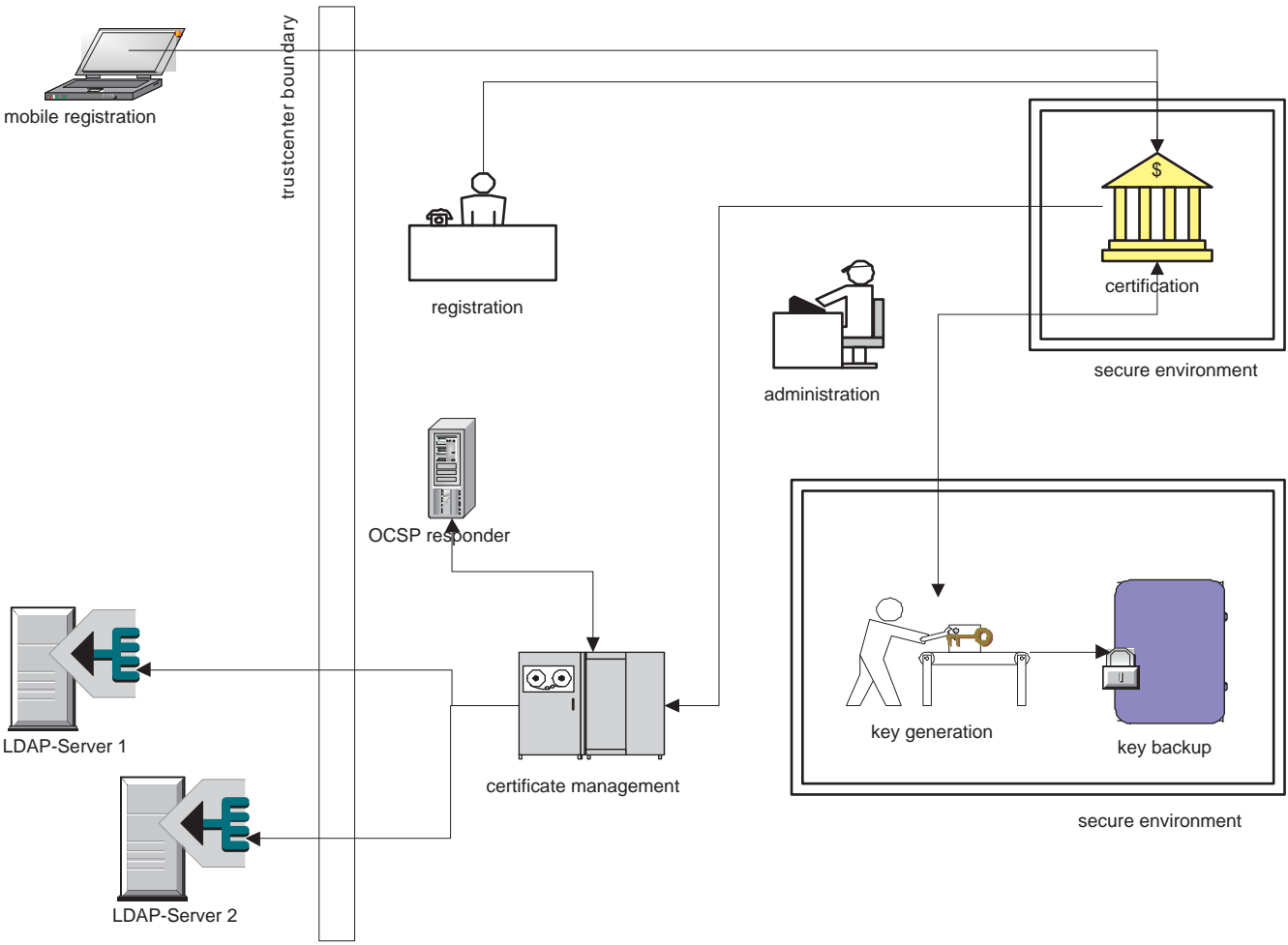

mobile registration

trustcenter boundary

registration

administration

certification

secure environment

OCSP responder

LDAP-Server 1

LDAP-Server 2

certificate management

key generation

key backup

secure environment